\documentclass[traditabstract
,letter
]{aa}  
\usepackage{graphicx,amsmath}
\usepackage{txfonts}

\begin{document}
\title{Temporal variations in the evaporating atmosphere of the exoplanet HD\,189733b}

\author{
A.~Lecavelier des Etangs\inst{1,2}
\and
V.~Bourrier\inst{1,2}
\and
P.~J.~Wheatley\inst{3}
\and
H.~Dupuy\inst{1,2}
\and
D.~Ehrenreich\inst{4}
\and
A.~Vidal-Madjar\inst{1,2}       
\and
G.~H\'ebrard\inst{1,2}       
\and
G.~E.~Ballester\inst{5}
\and
J.-M.~D\'esert\inst{6}       
\and
R.~Ferlet\inst{1,2}       
\and
D.~K.~Sing\inst{7}      
 }
   
\titlerunning{Temporal variations in the atmosphere of the exoplanet HD\,189733b}


   \institute{
   CNRS, UMR 7095, 
   Institut d'astrophysique de Paris, 
   98$^{\rm bis}$ boulevard Arago, F-75014 Paris, France
   \and
   UPMC Univ. Paris 6, UMR 7095,
   Institut d'Astrophysique de Paris, 
   98$^{\rm bis}$ boulevard Arago, F-75014 Paris, France
   \and
   Department of Physics, University of Warwick, Coventry CV4 7AL, UK
   \and
   UJF-Grenoble 1 / CNRS-INSU,  
   Institut de Plan\'etologie et d'Astrophysique de Grenoble (IPAG) UMR 5274,
   Grenoble, France 
   \and
   Lunar and Planetary Laboratory, University of Arizona, 1541 E. University Blvd., Tucson, AZ 85721-0063, USA
    \and
    Harvard-Smithsonian Center for Astrophysics, 60 Garden Street, Cambridge, MA 02138, USA
    \and
    Astrophysics Group, School of Physics, University of Exeter, Stocker Road, Exeter EX4 4QL, UK
   }
   
   \date{} 
 
  \abstract{
Atmospheric escape has been detected from the exoplanet \object{HD\,209458b} through 
transit observations of the hydrogen Lyman-$\alpha$ line. 
Here we present spectrally resolved Lyman-$\alpha$ transit observations of 
the exoplanet
\object{HD\,189733b} at two different epochs. These 
HST/STIS
observations show for the first time, that there are 
significant temporal variations in the physical conditions of an evaporating 
planetary atmosphere. 
While atmospheric hydrogen is not detected in the first epoch observations, it is observed at 
the second epoch, producing a transit absorption depth of 14.4$\pm$3.6\% between velocities of 
-230 to -140\,km\,s$^{-1}$. Contrary to HD\,209458b, these high velocities cannot arise from 
radiation pressure alone and require an additional acceleration mechanism, such as 
interactions with stellar wind protons. The observed absorption can be explained by an 
atmospheric escape rate of neutral hydrogen atoms of about 10$^9$\,g\,s$^{-1}$, 
a stellar wind with a velocity of 190\,km\,s$^{-1}$ and a temperature of $\sim$10$^5$\,K. 

An X-ray flare from the active star seen 
with Swift/XRT
8~hours before the second-epoch observation 
supports the idea that the observed changes within the upper atmosphere 
of the planet can be caused 
by variations in the stellar wind properties, or by variations in the stellar energy input to 
the planetary escaping gas (or a mix of the two effects). 
These observations provide the first indication of interaction 
between the exoplanet's atmosphere and stellar variations.

%
}


   \maketitle
%

\section{Introduction}
 \label{Introduction}

Observations of the transiting extrasolar planet \object{HD\,209458b} 
in the Lyman-$\alpha$ line of atomic hydrogen (\ion{H}{i}) have revealed 
that this planet is losing gas (Vidal-Madjar et al.\ 2003). 
Subsequent theoretical studies indicate that atmospheric escape (so-called `evaporation')
arises from the intense stellar
X-ray and extreme ultraviolet energy input into the upper atmosphere (Lammer et al.\ 2003; 
Lecavelier des Etangs et al.\ 2004; Yelle 2004), leading to moderate 
escape rates 
for massive hot-Jupiters, or to formation of planetary remnants when strong evaporation 
implies a dramatic change in the planet mass (Lecavelier des Etangs et al.\ 2004, 2007; 
Charpinet et al.\ 2011). 

Despite the importance of evaporation on the fate of planets at 
short orbital distances, the physics of the exospheric gas and role of the star-planet system 
properties remain debated (Garc\'ia Mun\~oz 2007; Schneiter et al.\ 2007; Holmstr\"om et al.\ 
2008; Lecavelier des Etangs et al.\ 2008; Murray-Clay et al.\ 2009; Ben-Jaffel \& Sona 
Hosseini 2010; Guo 2011). This is further compounded by the limited number of observations 
(Vidal-Madjar et al.\ 2004; Ballester et al.\ 2007; Ehrenreich et al.\ 2008; Fossati et al.\ 
2010; Linsky et al.\ 2010), which include non-spectrally resolved Lyman-$\alpha$ 
observations of the exoplanet \object{HD\,189733b} (Lecavelier des Etangs et al.\ 2010). \\


\section{Observations, data analysis, and results}

\subsection{Observations}

To address these problems, we observed two transits of HD\,189733b on 6~April 2010 and 
7~September 2011 with the Space Telescope Imaging Spectrograph (STIS) onboard the Hubble 
Space Telescope (HST). The data consist of time-tagged observations obtained with the G140M 
grating, producing time-resolved spectra from 1195 to 1248\,\AA\ with a spectral resolution of 
about 20\,km\,s$^{-1}$ at 1215.6\,\AA\ (Lyman-$\alpha$) with exposure times of 1800 to 2100~seconds. 
Between each consecutive
HST orbit, data acquisition is interrupted for about 3500~seconds by the Earth's 
occultation. The obtained spectra show stellar emission lines of 
\ion{H}{i} Lyman-$\alpha$, \ion{Si}{iii} (1206.5\,\AA), \ion{O}{v} (1218.3\,\AA) 
and the \ion{N}{v} doublet (1242.8\,\AA\ and 1238.8\,\AA). 

\begin{figure*}[tb]
\hbox{
\includegraphics[angle=-90,width=\columnwidth]
{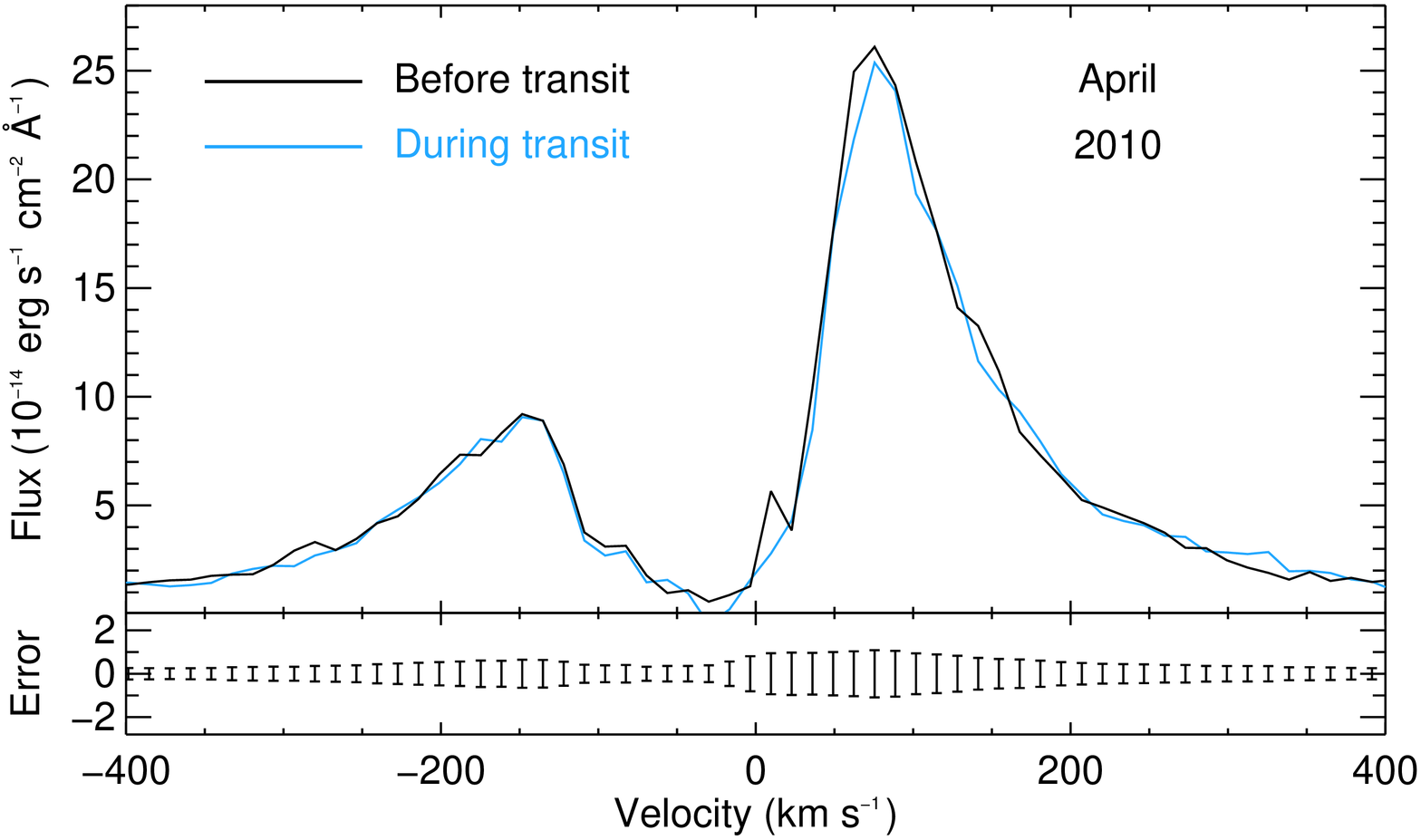}
\includegraphics[angle=-90,width=\columnwidth]
{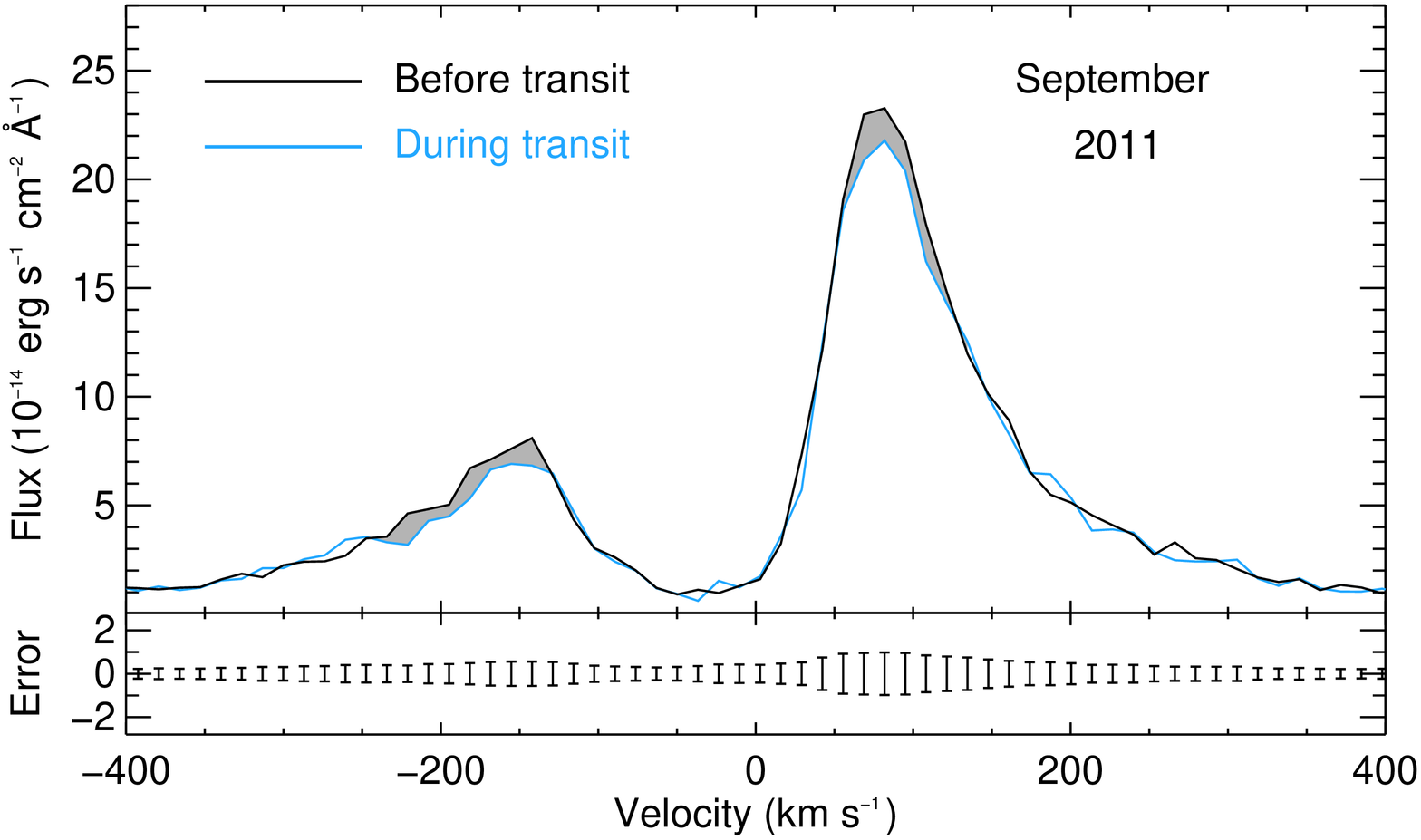}
}
\caption[]{Lyman-$\alpha$ emission line of HD\,189733 in April 2010 and September 2011. Spectra 
obtained before (black) and during the transits (blue) are displayed as a function of radial velocity
relative to the star. The double-peaked profile results from a single stellar emission line absorbed 
at the center by interstellar hydrogen, which produces a broad absorption feature from about -100 to 
+50\,km\,s$^{-1}$. While no transit signatures are detected in 2010, two absorption regions are 
detected at more than \hbox{3-$\sigma$} during the transit of 2011; these regions are plotted by gray zones. 
They are seen at the top of the red wing around +80\,km\,s$^{-1}$ and, most significantly, 
in the blue wing with a $\sim$100\,km\,s$^{-1}$ wide absorption around -200\,km\,s$^{-1}$.
}
\label{Lyman-alpha spectra}
\end{figure*}

For both the 2010 and 2011 observations, data were recorded during two HST orbits before 
the transit, one orbit during the transit, and one orbit after the transit. We measured 
the transit signature of the planetary atmosphere by comparing spectra taken during 
transit to those taken before and after the transit event. For each stellar line, we 
calculated transit light curves of the total emission flux and of the flux within given 
wavelength ranges; the signature of the planetary atmosphere is detected as an excess 
absorption during the planet's transit. Because no atmospheric signature is detected in 
the \ion{Si}{iii}, \ion{O}{v}, or \ion{N}{v} lines, 
hereafter we consider only the Lyman-$\alpha$ observations. 

The Lyman-$\alpha$ 
line is the brightest stellar line in our STIS spectra from 1195 to 1248\,\AA, and has a 
total flux of about 1.8$\times$10$^{-13}$\,erg\,s$^{-1}$\,cm$^{-2}$ 
(about 10 times brighter than for \object{HD\,209458}). 
With the resulting high signal-to-noise ratio, there is no need to co-add several 
observations to detect the signature of the atmosphere, which was necessary for the STIS observations 
of HD\,209458b and for the observations of HD\,189733b with the Advanced Camera for 
Surveys (ACS) of the HST (Vidal-Madjar et al.\ 2003; Lecavelier des Etangs et al.\ 2010). 
This allows for the first time a search for temporal 
variations in the physical conditions of the planetary upper atmosphere between two 
observational epochs (here 17 months apart), as performed for the deeper atmosphere using 
emission spectroscopy (Grillmair et al.\ 2008).

\subsection{Detection of temporal variations in the evaporating atmosphere}

The Lyman-$\alpha$ emission line from \object{HD\,189733} is spectrally resolved. At the resolution of the 
G140M grating, the line is composed of two peaks separated by a deep absorption due 
to interstellar atomic hydrogen (Fig.~\ref{Lyman-alpha spectra}). 
In the raw data, the stellar emission line is superimposed with the geo-coronal 
airglow emission from the upper atmosphere of the Earth (Vidal-Madjar et al.\ 2003). 
This geo-coronal emission can be well estimated and removed in the final spectrum 
using the CALSTIS data pipeline (version 2.32 of 5 November 2010). Independent 
re-analysis of raw data using the same methodology as for STIS observations of 
HD\,209458 (D\'esert et al.\ 2004) confirms that the airglow does not affect our 
measurements. Moreover, because we used a narrow spectrograph slit of 0.1\arcsec, the 
airglow contamination is limited to the central part of the Lyman-$\alpha$ line and 
does not contaminate the line wings where the transit atmospheric signatures are 
detected (see below). 
The data of September 2011 exhibit a notably low airglow emission level. 

The Lyman-$\alpha$ observations of April 2010 do not show 
an atmospheric transit signature. In these data, the transit depth for the total flux of 
the whole Lyman-$\alpha$ line is 2.9$\pm$1.4\%, which agrees with the 2.4\% transit depth 
of the planet body alone as seen from the visible to the near-infrared 
(D\'esert et al.\ 2009, 2011; Sing et al.\ 2011). In addition, 
we see no excess absorption in any portion of the Lyman-$\alpha$ spectral line profile. 

However, this situation strongly contrasts with the observations of September 2011, 
in which we see an excess absorption in the total flux of the Lyman-$\alpha$ line 
that yields a transit depth of 5.0$\pm$1.3\%. 
This level is consistent with the results obtained with the non-resolved 
HST/ACS spectra of 2007-2008 (Lecavelier des Etangs et al.\ 2010). 
Importantly, the line profile shows two deep absorption regions at specific wavelength intervals 
during the transit: 
first in the blue part of the spectrum from about -230\,km\,s$^{-1}$ to -140\,km\,s$^{-1}$, 
and in the red part of the spectrum from 60 to 110\,km\,s$^{-1}$. 

In the blue wing of the Lyman-$\alpha$ spectrum of 2011, the most significant absorption is 
visible in the range of \hbox{-230} to \hbox{-140\,km\,s$^{-1}$}, 
which gives an absorption depth of 14.4$\pm$3.6\% 
(\hbox{4-$\sigma$} detection) corresponding to an excess absorption due to hydrogen atoms of 12.3$\pm$3.6\%. 
The false-positive probability to find such an excess anywhere in the whole 
searched range of -350 to -50\,km\,s$^{-1}$ is only 7\%. 
In the red wing of the Lyman-$\alpha$ spectrum 
of 2011, absorption is found in the range between 60 to 110\,km\,s$^{-1}$, which yields 
an absorption depth of 7.7 $\pm$ 2.7\% (\hbox{3-$\sigma$} detection) 
and corresponds to an excess absorption 
due to hydrogen atoms of 5.5$\pm$2.7\%. The false-positive probability to find such an 
excess over the whole searched range [40 to 200\,km\,s$^{-1}$] is 39\%. This high probability 
shows that the absorption seen in the red wing of the spectrum may not be real and is 
possibly caused by statistical noise in the data, although interestingly enough, a similar 
absorption was also observed in HD\,209458b (Vidal-Madjar et al.\ 2003). 

\begin{figure}[tb]
\hbox{
\includegraphics[angle=-90,width=\columnwidth]
{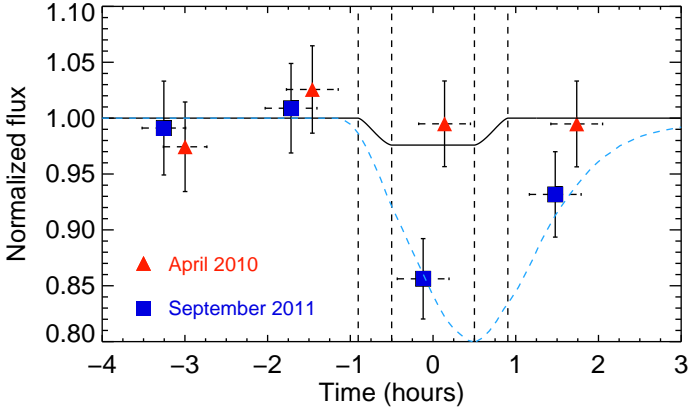}
}
\caption[]{
Plot of the flux between -230 and -140\,km\,s$^{-1}$ in the blue wing of the 
Lyman-$\alpha$ line as a function of time relative to the center of the planetary transit.
Vertical dashed lines 
show the beginning and end of ingress and egress of the transit. The red triangular 
symbols are for the 2010 observations, while the blue square symbols correspond to 
observations of 2011. Horizontal error bars centered on the symbols show the duration of 
the exposures in each HST orbit. 
The time-tagged data allow independent sub-exposures to 
be extracted within each HST orbit (not shown here), 
resulting in the same transit signal within error bars.
The light curve of the planet's transit at optical 
wavelengths is displayed as a solid black line. The blue dashed line shows the calculated 
flux using the numerical simulation with an EUV ionizing flux 5~times the solar value, a 
stellar wind of protons with a temperature $T$$\sim$10$^5$\,K, a velocity 
$v$$\sim$190\,km\,s$^{-1}$ and density $n$$\sim$3$\times$10$^3$\,cm$^{-3}$ together with an 
atmospheric escape rate of 10$^9$\,g\,s$^{-1}$.
}
\label{flux vs time}
\end{figure}

None of the excess absorption features detected in the September 2011 data are seen in 
the April 2010 data. We conclude that there are significant temporal variations of the 
physical conditions within the extended exosphere of this extrasolar planet between these 
two epochs (Fig.~\ref{flux vs time}). 

\subsection{Models}

In September 2011, the absorption depth and velocity range show that neutral hydrogen atoms are 
present up to very high altitudes at velocities exceeding the escape velocity 
of 60\,km\,s$^{-1}$; this unambiguously demonstrates that atmospheric gas must be escaping 
from HD\,189733b. For HD\,209458b, the \hbox{Lyman-$\alpha$} excess absorption was 
observed in the spectral range between \hbox{-130} to -50\,km\,s$^{-1}$, which is readily 
explained by the stellar radiation pressure accelerating hydrogen atoms up to 
\hbox{-130}\,km\,s$^{-1}$ 
(Lecavelier des Etangs et al.\ 2008). The case here of HD\,189733b, which shows excess 
absorption at higher velocities between -230 and -140\,km\,s$^{-1}$, is 
more challenging to explain. Assuming a distance of 19.3~pc and following the method 
described in Ehrenreich et al.\ (2011), we estimated the interstellar medium absorption 
and calculated 
the Lyman-$\alpha$ emission line profile as seen from the planet. From the 
extinction-corrected line profile we estimated that radiation pressure can accelerate 
hydrogen atoms up to a radial velocity of -120\,km\,s$^{-1}$ in the exosphere of this planet 
(below this radial velocity the stellar flux at the corresponding wavelength in the core 
of the emission line is sufficiently high for the radiation pressure to exceed the stellar 
gravity). Therefore, an additional acceleration mechanism beyond radiation pressure is 
required to explain the high radial velocities of hydrogen measured during the transit. 
Charge exchange with stellar wind protons can produce the observed high velocities 
(Holstr\"om et al.\ 2008; Ekenb\"ack et al.\ 2010). 

\begin{figure}[tb]
\hbox{
\includegraphics[angle=-90,width=\columnwidth]
{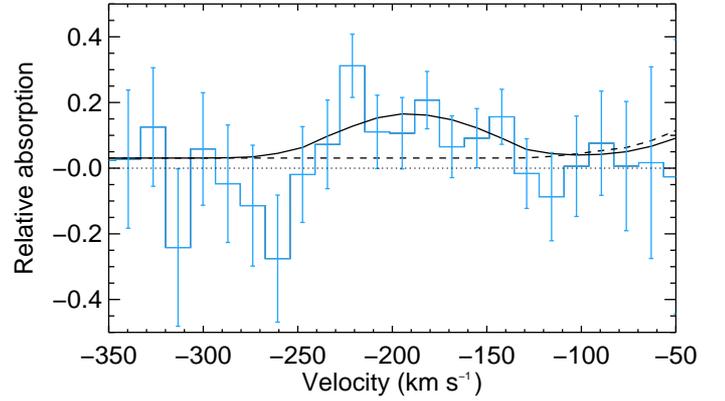}
}
\caption[]{Plot of the relative absorption observed in the blue wing of the 
Lyman-$\alpha$ stellar line (blue histogram) for the transit of September 7, 2011. The 
dashed line shows the model with radiation pressure only; in this case, there is no 
absorption at radial velocities below $\sim$-120\,km\,s$^{-1}$, resulting in a large 
$\chi^2$ of 22.8. If a stellar wind and charge exchange is considered, hydrogen atoms can 
be accelerated to the higher observed velocities. The model with an escape rate of 
10$^9$\,g\,s$^{-1}$ (solid line) gives a $\chi^2$ of 13.0 for 17 degrees of freedom. 
}
\label{Absorption vs velocity}
\end{figure}

To investigate this possibility and interpret the observed H{\sc i} light curve, we 
developed a numerical Monte-Carlo simulation of the hydrogen atom dynamics. 
The details of the model will be given in a forthcoming paper 
(Bourrier et al.\ in preparation). In this N-body simulation, hydrogen atoms are 
released from HD\,189733b's upper atmosphere and 
are rapidly accelerated by the radiation pressure up to 120\,km\,s$^{-1}$ and then to higher 
velocities by charge exchange with protons from the stellar wind. This dynamical model 
provides the time-dependent distribution of positions and velocities of the escaping 
hydrogen atoms in the cloud surrounding HD\,189733b. With this information, we calculated 
the corresponding absorption over the stellar emission line and the resulting transit 
light curve, which can be directly compared with the observations. We find that the 
observations are well-fitted with an escape rate of neutral hydrogen of 
about 10$^9$\,g\,s$^{-1}$ 
and a stellar wind with a temperature $T$$\sim$10$^5$\,K, 
a density $n$$\sim$3$\times$10$^{3}$\,cm$^{-3}$, and a velocity $\sim$190\,km\,s$^{-1}$. 
The best-fit model yields a $\chi^2$ of 13.0 for 17~degrees of freedom 
for the absorption spectrum given in Fig.~\ref{Absorption vs velocity}. 
The EUV flux controlling the neutral hydrogen ionizing timescale should be about 
5~times the solar value to explain the moderate absorption observed after the transit of 
the planet (Fig.~\ref{flux vs time}). 

\begin{figure*}[tb]
\includegraphics[angle=0,width=0.78\textwidth]{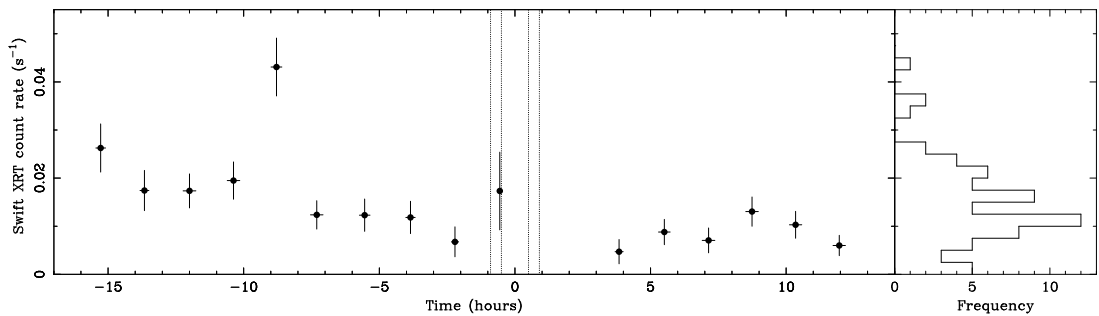}
\caption[]{Swift XRT X-ray light curve of HD\,189733 about the time of the September 7, 
2011 transit. 
The data were binned into one point per snapshot visit, with typical 
exposure times of about 27~minutes. Vertical dashed lines show the beginning and end of 
ingress and egress of the transit. A bright flare occurred about 8 hours before the 
transit. The observed average count rate during the flare snapshot is a factor 3.6 higher 
than the mean for the whole light curve, indicating a peak X-ray flux (0.3-3\,keV) that must be less than 1.3$\times$10$^{-12}$\,erg\,s$^{-1}$\,cm$^{-2}$. For comparison, 
the right panel shows the distribution for 63~epochs of Swift measurements 
(including the 16 obtained in September 2011), covering a wide range of timescales. 
The flare occurring shortly before 
the transit is the highest X-ray flux of all 63~measurements.
}
\label{X-ray}
\end{figure*}

\section{Swift X-ray simultaneous observations}

The evaporation of hot Jupiters is driven by the X-ray/EUV irradiation of the 
planet by its parent star. To quantify the level of X-ray 
irradiation at the time of our observations in September 2011, we obtained 
contemporaneous observations with the X-ray telescope (XRT) of the Swift 
spacecraft (Gehrels et al.\ 2004). 
A source was detected toward HD\,189733 with a mean count 
rate of 0.0119$\pm$0.0007\,s$^{-1}$;  
consistent with previous XMM-Newton observations (Pillitteri et al.\ 2010), 
we did not detect
the M-star binary companion, \object{HD\,189733B}, but we did find evidence for a very weak
hard X-ray source located about 13\arcsec\ south of \object{HD\,189733A}.

The Swift XRT spectrum of HD\,189733 can be fitted with a multi-temperature optically-thin 
thermal plasma model (Mewe et al.\ 1985; Liedahl et al.\ 1995) that is 
typical of the coronal X-ray emission from active stars. Using a three-temperature 
fit (temperatures of 0.12, 0.46 and 4.5\,keV) we found an average X-ray flux 
in the 0.3-3\,keV band of 3.6$\times$10$^{-13}$\,erg\,s$^{-1}$\,cm$^{-2}$, 
consistent with
XMM-Newton observations of HD\,189733 (Pillitteri et al.\ 2011). 
This flux corresponds to a 
planetary irradiation rate of 1.2$\times$10$^{24}$\,erg\,s$^{-1}$, 
which could drive a mass loss rate of up to 1.0$\times$10$^{11}$\,g\,s$^{-1}$ 
(assuming 100\% evaporation efficiency and taking evaporation enhancement by tidal 
forcing into account; Lecavelier des Etangs et al.\ 2007; Erkaev et al.\ 2009). 
Assuming a realistic emission measure distribution (Sanz-Forcada et al.\ 2011), 
we can estimate the total luminosity across the X-ray/EUV band at the time of our 
observation to be 7.1$\times$10$^{28}$\,erg\,s$^{-1}$, 
corresponding to an energy-limited evaporation rate of 
4.4$\times$10$^{11}$\,g\,s$^{-1}$. 
The X-ray irradiation is consistent with the estimated 
escape rate, which would thus require around 1\% efficiency
in the conversion of input energy into mass loss (Ehrenreich \& D\'esert 2011). 
But this is only a lower limit because 
the estimated escape rate of neutral hydrogen
atoms of about 10$^{9}$\,g\,s$^{-1}$ does not include the escape of ionized hydrogen 
at the exobase of the atmosphere, and it is therefore a lower limit for the net
escape from HD189733b.

The Swift XRT light curve of HD\,189733 (Fig.~\ref{X-ray}) shows that the star exhibits 
significant X-ray variability, and most notably, a bright flare that occurred 
about 8~hours before the planetary transit. This flare could explain the 
observed variations in the extended cloud of high-velocity hydrogen atoms 
escaping the planet, because this could affect the properties of the stellar wind 
needed to accelerate the atoms to the observed radial velocities. Besides, 
the enhanced X-ray/EUV irradiation associated with this flare 
must lead to a significantly enhanced escape rate. With our best-model 
parameters, an enhanced escape rate leads to a more extended exospheric hydrogen 
cloud and thus to a stronger absorption after about one hour; the absorption 
remains at a high level during a typical ionization timescale, which is 
constrained from the post-transit Lyman-$\alpha$ observations to be about 5 hours. 
Therefore, an X-ray flare occurring 8~hours earlier is expected to lead to 
higher escape rate that is then detectable in Lyman-$\alpha$. 

\section{Conclusion}

Whether they are related to the observed X-ray flare or not, the 
temporal variations in the 
evaporating atmosphere of HD\,189733b are clearly detected in Lyman-$\alpha$. 
The variability of the neutral hydrogen cloud around HD\,189733b can explain the 
high dispersion of absorption depth measurements in spectrally non-resolved 
Lyman-$\alpha$ observations (Lecavelier des Etangs et al.\ 2010); 
combining this with the present high 
signal-to-noise ratio spectrally resolved observations, we conclude that 
escape signatures are detected in about half of the total five transits 
observed in Lyman-$\alpha$. More simultaneous X-ray and Lyman-$\alpha$ observations 
are needed to obtain a better picture of the complex relationship between the 
stellar energetic input to the planet and the atmosphere's response to it, 
and to constrain theoretical  
models of a space weather event on hot-Jupiters ({\it e.g.} Cohen et al.\ 2011). 
The HD\,189733 system appears to be the target of choice, but future observations 
should also enlarge the diversity of stellar and planetary system properties 
to better distinguish the effects of the stellar-planet interactions from the 
intrinsic variability in the observed atmospheres.

\begin{acknowledgements}

Based on observations made with the NASA/ESA Hubble Space Telescope, obtained at the Space Telescope Science Institute, which is operated by the Association of Universities for Research in Astronomy, Inc., under NASA contract NAS 5-26555.
This research has made use of data obtained from NASA's Swift satellite. 
G.E.B.\ acknowledges financial support by this program through STScI grant 
HST-GO-11673.01-A to the University of Arizona.
These observations are associated with program \#11673. This work has been supported by an award from the {\it Fondation Simone et Cino Del Duca}. 

\end{acknowledgements}

\end{document}